\documentclass{jps-cp}
\usepackage{txfonts} 

\usepackage{braket}
\usepackage{here}
\usepackage{comment}
\usepackage{color}
\title{Transverse Field Dependence of the Ground State in the $Z_2$ Bose-Hubbard Model}

\author{Yuma \textsc{Watanabe}$^{1}$, Shohei \textsc{Watabe}$^{1,2}$, and Tetsuro \textsc{Nikuni}$^{1}$}

\inst{$^{1}$Department of Physics, Tokyo University of Science, Shinjuku, Tokyo,162-8601, Japan \\
$^{2}$Faculty of Engineering, Computer Science and Engineering, Shibaura Institute of Technology 3-7-5 Toyosu, Koto-ku, Tokyo 135-8548, Japan}

\email{1221712@ed.tus.ac.jp}

\recdate{July 7, 2022}

\abst{The study of interaction between the particle and lattice degrees of freedom is one of the central interests in the quantum many-body systems.
The $Z_2$ Bose-Hubbard model has been proposed to describe ultracold bosons in a dynamical optical lattice. 
This model introduces the lattice degrees of freedom by placing half-spins on the bonds between neighboring lattice sites.
In this study, we investigate the effect of spin fluctuations on the ground state by using the density-matrix renormalization group method.
By calculating the spin structure factor and the compressibility, we show that there is a phase transition between two spatially nonuniform states.
We also discuss the ground state in the strong transverse magnetic limit.}

\kword{Optical lattice, Ultracold gas, Lattice gauge theory, Phase transition}

\begin{document}
\maketitle

\section{Introduction}
	Ultracold gases in an optical lattice provide a tremendous platform for studying the strongly-correlated many-body systems.
	In particular, ultracold Bose gases in an optical lattice are described by the Bose-Hubbard model, and its controllability allows experimental access to observe the Superfluid--Mott-insulator transition \cite{format4,format5}.
	Due to the high controllability, a variety of the optical lattice system prohibiting novel and interesting phenomena, such as the topological orders and the supersolids, have been broadly studied \cite{format8}.
	Nonetheless, the studies of the ultracold Bose gases in dynamical lattices have still been limited compared to the static cases.
	
	Understanding the role of the interactions between particles and lattice degrees of freedom is a crucial issue in quantum many-body physics.
	Fermion-phonon interactions have been extensively studied, which led to important findings such as the elucidation of the mechanism of superconductivity.
	On the other hand, similar problems for bosonic systems have not been studied compared to fermionic systems.	

	With these backgrounds, the $Z_2$ Bose-Hubbard model has been introduced to describe ultracold bosons in a dynamical optical lattice \cite{format1}.
	In this model, the lattice degree of freedom is introduced through the $Z_2$ gauge field, which is equivalent to placing spins on the bonds between neighboring lattice sites in the Bose-Hubbard model.
	Despite the absence of a Fermi surface, it has been found that this model exhibits a bosonic analog of Peierls phase transition, that is, a structural phase transition accompanied by the opening of the energy gap \cite{format1}. 
	Furthermore, this model leads to vigorous studies on the symmetry-protected topological phase \cite{format2}, the topological Peierls insulator, and the Peierls supersolid \cite{format3}.
	
	In this paper, we study the ground state of the $Z_2$ Bose-Hubbard model by using the density-matrix renormalization group (DMRG) method \cite{format6, format7}.	
	In particular, we investigate the effect of the transverse field on the ground state.
	We show the existence of a phase transition between two different translational symmetry broken states.
	The model undergoes a transition between the compressible and incompressible phases with the increasing transverse field.

\section{$Z_2$ Bose-Hubbard model}
The $Z_2$ Bose-Hubbard Hamiltonian is given by
\begin{align}
	\hat{H}&= -J\sum_i (\hat{b}_i^\dagger\hat{b}_{i+1}+\hat{b}_{i+1}^\dagger\hat{b}_i)+\frac{U}{2}\sum_i\hat{n}_i(\hat{n}_i-1)-\mu\sum_i\hat{n}_i \nonumber\ \\
	&\hspace{1cm} -\alpha\sum_i (\hat{b}_i^\dagger\hat{\sigma}_i^z\hat{b}_{i+1}+\hat{b}_{i+1}^\dagger\hat{\sigma}_i^z\hat{b}_i)+\frac{\Delta}{2}\sum_i\hat{\sigma}_i^z+\beta\sum_i\hat{\sigma}_{i}^x,
\end{align}
where $\hat{b}_i (\hat{b}^\dagger_i)$ is the annihilation (creation) operator that acts on the site $i$, and $\hat{n}_i=\hat{b}_i^\dagger\hat{b}_i$ is the particle number operator.
$\sigma^z_i$ and $\sigma^x_i$ are Pauli operators accompanied by spin-1/2 and act on a bond between neighboring sites.
The first three terms represent the conventional Bose-Hubbard Hamiltonian, which includes the hopping term, the on-site inter-atomic interaction term, and the chemical potential term. 
The fourth term represents the lattice-dependent tunneling between neighboring sites.
The effective hopping $J^\prime = -J-\alpha\braket{\hat{\sigma}_i^z}, (\alpha > 0)$ is maximized when the spins on the bonds are in the up state and minimized when they are in the down state.
 The last two terms represent the spin dynamics, where $\Delta$ and $\beta$ are the strengths of the longitudinal and the transverse fields. 
 In the DMRG calculation in this paper, we set the total lattice number and the particle number cutoff to $L=30$ and $n_0 = 2$ respectively, and the maximum bond dimension is truncated to $\chi=40$.
 Hereafter, we fix the value of parameters to $U=10$, $\alpha=0.5$, and $\Delta=0.85$ respectively, and take $J=1$ to set the energy scale.

When the transverse field $\beta$ is small compared to the hopping ($J \gg \beta= 0.02$), the ground state is determined by the competition between the strength of the lattice-dependent hopping $\alpha$ and the spin-flip energy $\Delta$.
When the spin-flip energy is much larger than the lattice-dependent hopping ($\Delta\gg\alpha$), the spin prefers the down-state, the hoping effect is minimum, and the ground state is a Mott insulator state.
On the other hand, when the lattice-dependent hopping is sufficiently large ($\alpha\gg\Delta$), all the spins are in the up-state, and the hopping effect is maximum.
Therefore, the state is in the superfluid state.
When the two parameters, lattice-dependent hopping and energy difference, are comparable ($\Delta\sim\alpha$), the translational symmetry is broken, and the states have spatial structures.
The previous research \cite{format1} revealed that the system shows two types of translational-symmetry broken states called the commensurate/incommensurate bond order wave (cBOW/iBOW).
The cBOW exhibits spatial modulation at integer wavelengths, whereas the iBOW has non-integer wavelengths.

\begin{table}[htb]
\caption{The physical features of the four different ground state.}
\label{t1}
\begin{tabular}{c|cc}
\hline
 & Peak of the spin structure factor & Compressibility \\
 \hline
SF & $=0$ & $\neq 0$ \\
MI & $=0$ & $=0$\\
cBOW & $\neq 0$ & $=0$\\
iBOW & $\neq 0$ & $\neq 0$ \\
\hline
\end{tabular}
\end{table}

These four states, i.e., SF, MI, cBOW, and iBOW, can be completely distinguished by two physical quantities: the spin structure factor $S(k)$ and the compressibility $\kappa$.
The spin structure factor is given by
	\begin{equation}
	S(k) = \frac{1}{L^2}\sum_{i, j}e^{i(x_i - x_j)k}\Braket{(\hat{\sigma}_i^z-\bar{\sigma^z})(\hat{\sigma}_j^z-\bar{\sigma}^z)},
	\end{equation}
where $\bar{\sigma}^z = \sum_i \braket{\hat{\sigma}_i^z}/L$.
The compressibility is given by $\kappa = \frac{\partial \rho}{\partial\mu}$.
The spin structure factor has peaks at a certain wavenumber $k_0$ when the ground state has spatial modulations, and its heights can be used as the order parameter of the structural transition.
The wavelength can be calculated as $\lambda = 2\pi/k_0$ since the contribution of the sharp peak of the spin structure factor is dominant to the ground state modulation.
The compressibility is also useful for distinguishing the four states. 
The previous study \cite{format1} has revealed that the Mott-insulator and the cBOW are incompressible ($\kappa=0$), while the superfluid and the iBOW are compressible($\kappa\neq 0$). 
Table\ref{t1} summarizes these characteristic features.

\section{Dependence on the Transverse Field and Incommensurate--Commensurate Bond Order Wave Phase Transition}
In this section, we investigate the effect of the spin fluctuation on the ground state of the $Z_2$ Bose-Hubbard model. 
We first study how the spatial structure changes with the increasing transverse field by calculating the peak values of the spin structure factor. 
We next calculate the compressibility to confirm the iBOW--cBOW transition.
Finally, we discuss the features of the ground state in the limit of the strong transverse fields.
		\begin{figure}[H]
		\centering
		\includegraphics[scale=0.9]{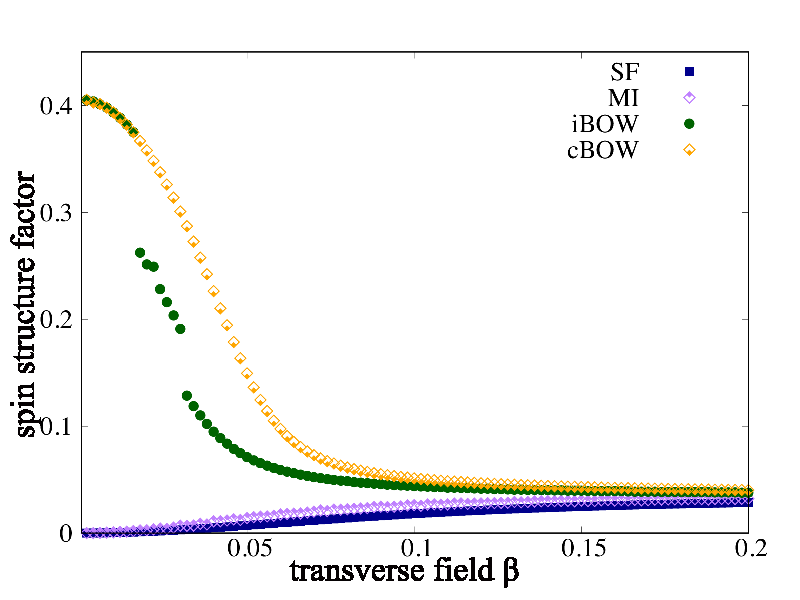}
		\caption{The peak values of the spin structure factor as a function of the transverse field. We calculate the spin structure factor in the superfluid (SF, blue-triangle), the Mott-insulator (MI, purple-diamond), and the commensurate and incommensurate bond order wave states (cBOW, iBOW, orange-square and green-circle, respectively). 
		}
		\label{f1}
		\end{figure}
	Figure \ref{f1} shows the transverse field dependence of the peak of the spin structure factor in the case of the superfluid (blue-triangle), Mott-insulator(purple-diamond), and commensurate (orange-square) and incommensurate (green-circle) bond order wave states.
	The region where the transverse field is close to zero ($\beta < 0.005$) is excluded from this plot.
	This is because in this region, the transverse field is not large enough to allow spins to move and flip freely, and thus the ground state strongly depends on the initial state, making it difficult to obtain the true ground state by the DMRG calculation.
	
	We use the value of the chemical potential $\mu = 0.50, 1.50, -0.40, -0.17$ for the calculation of SF, the MI, the cBOW, and the iBOW, respectively.
	In the case of the spatially uniform SF and MI states, one can see a gradual increase of the peak values of the spin structure factor toward specific values with increasing transverse field.
	In contrast, the peaks decrease significantly with the increasing the transverse field in the case of the bond order waves.
	This shows that for the sufficiently large transverse fields the spatial modulation of the ground state disappears and the spatial structure becomes uniform.
	Moreover, there are several points where the peak decreases discontinuously in the case of the iBOW.
		
		\begin{figure}[H]
		\centering
		\includegraphics[scale=0.3]{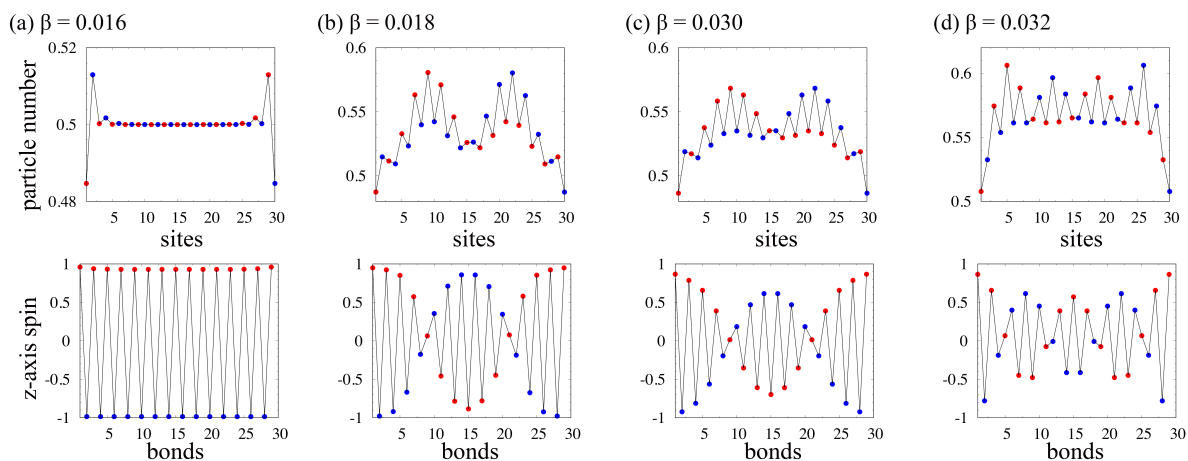}
		\caption{The particle number and configurations of $z$-axis spin just above and just below the two discontinuous points. These panels correspond to just below ($\beta=0.016$, (a)) and above ($\beta=0.018$, (b)) the first discontinuous point and just below ($\beta=0.030$, (c)) and just above ($\beta=0.032$, (d)) the second discontinuous point, respectively. These configurations show that the ground state has different spatial structures dependent on the magnitude of the transverse field.}
		\label{f2}
		\end{figure}

	To investigate the change in the ground states with the increasing transverse field, we calculate the spatial configuration of the local particle number $\braket{\hat{n}_i}$ and the $z$ component of the spin $\braket{\hat{\sigma}_i^z}$ at the discontinuous points of the peak of the spin structure factor in the case of the iBOW.
	The particle number and the spin just above and just below the two different discontinuous points are shown in Fig. \ref{f2}. 	
	
	Just below the first discontinuous point ((a) in Fig. \ref{f2}), the average particle density is $1/2$, and spins show the spatially discrete configuration taking values of $+1$ or $-1$.
	The previous study has shown that the ground state for the parameters considered here is the iBOW state when the transverse field is $\beta = 0.02$.
	When the transverse field is sufficiently small, the configurations of the particle number and the spin are similar to the cBOW state with the particle density $1/2$, showing spatial modulation with nearly integer value of the wavelength $\lambda\sim2.0$.
	In contrast, the particle density changes from $1/2$ to $16/30$ just above the first discontinuity (Fig. \ref{f2} (b)). 
	At this point, the configurations of the particle number and the spins are the same as in the iBOW reported in the previous study \cite{format2}.
	One ``extra" boson leads to the incommensurate density modulation with a long wavelength.
	The spins gradually change from $+1$ to $-1$ or vice versa, forming kinks in the spin configuration.
	
	After the first sharp decrease, the particle density remains at the constant value of $16/30$ until just below the second discontinuity (Fig. \ref{f2} (c)).
	The configurations of the particle number and spin are almost the same as just above the first discontinuity, except for the magnitude of $\braket{\hat{\sigma}_i^z}$. 
	Because the spins are tilted toward the $x$-axis, the $z$ component of spins are relatively small. 
	The particle density is changed from $16/30$ to $17/30$ just below the second discontinuous point (Fig. \ref{f2} (d)).
	In this case, there are two extra bosons, and the configurations become more complicated than just below the second discontinuity. The number of kinks increases from one to two, which is reflected in the particle number configuration, forming the two peaks. 
	\begin{figure}[H]	
		\centering
		\includegraphics[scale=0.8]{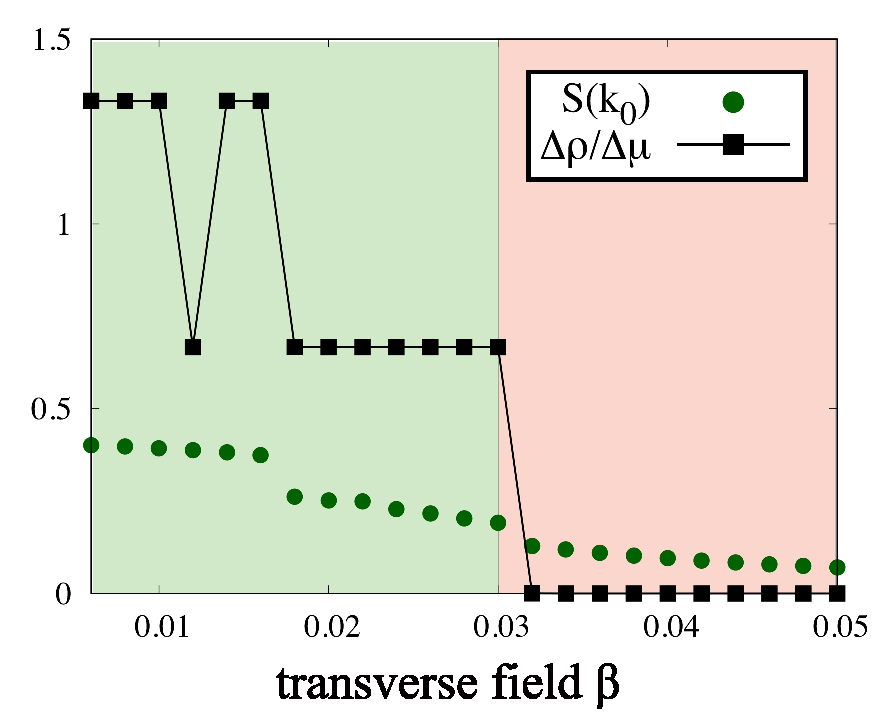}
		\caption{The peak of the spin structure factor (green dots) and the displacement of particle density in relation to the chemical potential $\Delta\rho/\Delta\mu \equiv [\rho(\mu+\delta\mu)-\rho(\mu)]/\Delta\mu$ (black dots-line) as a function of the transverse field.
		The peak of the spin structure factor remains finite as the transverse field increase.
		On the other hand, the value of $\Delta\rho/\Delta\mu $ suddenly drops to zero at the second discontinuous point.
		These indicates that there is a phase transition from the iBOW (light-green region) to the cBOW (orange region).}
		\label{f3}
	\end{figure}
	
	To more accurately confirm the existence of the phase transition, we calculate the rate of change of the particle density with respect to the chemical potential $\Delta\rho/\Delta\mu \equiv [\rho(\mu+\delta\mu)-\rho(\mu)]/\Delta\mu$ in the case of the iBOW.
	This quantity approximates the compressibility, which can be used to distinguish the different phases, as listed in Table \ref{t1}. 
	It should be noted that in the case of bond order waves, the particle density varies discontinuously with respect to the chemical potential, which makes the calculation of the compressibility numerically unstable.
	Figure \ref{f3} shows the peak of the spin structure factor $S(k_0)$ (green dots) and the value of $\Delta\rho/\Delta\mu $ (black dots-line) for increasing the transverse field. 
	The peak value in the spin structure factor keeps being finite, and it indicates the ground state has nonuniform spatial modulation. 
	However, the value of $\Delta\rho/\Delta\mu $ suddenly drops to zero from the non-zero value at the second discontinuous point. 
	Because the two bond order waves are characterized by the finite value of the peak in the spin structure factor and their compressibility, we conclude that the $Z_2$ Bose-Hubbard model undergoes the phase transition between the incommensurate (light-green region) and the commensurate (orange region) bond order waves with the transverse field increases.

 	We here note that the study of incommensurate states generally requires calculations on large systems since they are strictly speaking defined in infinitely lager systems.
	We recall that the previous study~\cite{format1} carried out scaling analysis and showed that the essential features of the spin structure factor and the compressibility do not strongly depend on the system sizes, and thus the physical properties of the incommensurate states are well captured in the finite-size calculations.
	Our results with $L=30$ are consistent with Ref.~\cite{format1}, and thus we conclude that the system size used in this study is sufficiently large to clearly see the phase transition between cBOW and iBOW.
			
	We finally comment on the ground state in the strong transverse field limit. 
	As the transverse field increases, the $z$ component of spin goes to zero, and the spins are oriented along the $x$-axis.
	Because in the $Z_2$ Bose-Hubbard Hamiltonian the particles are coupled only to the $z$ component of the spins, the contribution of the lattice-dependent hopping term vanishes in the limit of the strong transverse field. 
	Therefore, the ground state has no spatial modulation in this limit (see Fig. \ref{f1}). 
	When the spin is completely oriented along the $x$-axis, the contribution of the term containing $\hat{\sigma}^z_i$ vanishes, and the $Z_2$ Bose-Hubbard Hamiltonian becomes equivalent to the conventional Bose-Hubbard Hamiltonian.
		 
\section{Conclusion and Outlook}
In this paper, we have investigated the effect of spin fluctuation on the ground state in the $Z_2$ Bose-Hubbard model. 
By calculating the spin structure factor and  the rate of change of the particle density with respect to the chemical potential $\Delta\rho/\Delta\mu$, we have found that the phase transition from the iBOW to the cBOW occurs. 
In the iBOW phase, we showed the configurations of $\braket{\hat{n}_i}$ and $\braket{\hat{\sigma}_i^z}$ at two discontinuities in the peak value of the spin structure factor.
We also discussed the ground state in the strong transverse field limit and showed that the $Z_2$ Bose-Hubbard model becomes equivalent to the conventional Bose-Hubbard model. 
In future studies, it would be interesting to study the quench and sweep dynamics between the two different phases by varying the transverse fields. 

\section{Acknowledgment}
The authors thank M. Kunimi for his helpful comments and critical reading of the manuscript.
This work was supported by JST SPRING, Grant Number JPMJSP2151. 
S.W. was supported by JSPS KAKENHI Grant No. JP18K03499.

\end{document}